\documentclass[aps,prl,twocolumn,groupedaddress]{revtex4}

\usepackage{graphicx}
\usepackage{textcomp}
\usepackage{times}
\begin{document}


\title{Phase shift of a weak coherent beam induced by a single atom}

\author{Syed Abdullah Aljunid, Meng Khoon Tey, Brenda Chng,
 Timothy Liew, Gleb Maslennikov, Valerio Scarani and  Christian Kurtsiefer}
\email[]{christian.kurtsiefer@gmail.com}
\affiliation{Center for Quantum Technologies and Department of Physics,
  National University of Singapore, 3 Science Drive 2,  Singapore, 117543}
\date{\today}

\begin{abstract}
We report on a direct measurement of a phase shift on a weak coherent beam by
a single $^{87}$Rb atom in a Mach-Zehnder interferometer. A maximum
phase shift of about $1^\circ$  is observed experimentally. 
\end{abstract}


\maketitle

\section{Introduction\label{intro}}
While photons are the ideal carriers for transporting quantum information over
long distances, atoms can be used to store and process information. Thus, 
atom-photon interfaces will be important for implementing more
complex quantum information processing tasks~\cite{cirac:1997, duan:2001}.
The efficiency of information exchange between photonic `flying' qubits and
atoms or similar microscopic systems requires a strong interaction between
them, characterized e.g. by the scattering probability of a single photon. 
The traditional method to bring this probability close to
unity is to place an atom into high finesse cavity~\cite{rempe:2000,
kimble:2007}, where, in a simplified picture, a photon visits the atom
many times and hence increases its chance of being scattered. Recently,
however, it was shown that efficient scattering can also be achieved 
without cavity assistance by strong focusing, localizing the field
of the photon to a small region near the
scatterer~\cite{mk:2009,zumofen:2008}. A high scattering probability of
photons has been demonstrated experimentally for various microscopic
systems~\cite{imamoglu:2007,wrigge:2008,our_paper}. 

Apart from the power changing aspect of the scattering process, the presence
of the single atom in a focus of the light beam can also change its phase.
This may help to realize a photonic phase gate,
in which the phase of a photon is changed depending on the presence or the
internal state of the atom~\cite{walls:1990}. In such a scenario, the atom can
be viewed as a mediator for photon-photon interactions due to the non-linear
dispersion. This nonlinear phase shift has been investigated in experiments
involving cavities~\cite{turchette:1995,fushman:2008} and atomic 
ensembles~\cite{zibrov:1996}. It is interesting to perform a similar
experiment with a strongly focused optical mode, because of its much
reduced complexity compared to cavity QED experiments.

As a first step towards such an element, we report here on the direct
measurement of the phase shift the presence of a single $^{87}$Rb atom imposes
on a strongly focused coherent light field in a Mach-Zehnder
interferometer. There, the probe passes only once through the atom  
localization volume. Following~\cite{mk:2009,zumofen:2009}, a simple
theoretical model is used to describe the experimental results.

\section{Experimental setup\label{sec:setup}}
\begin{figure}
\begin{center}
\includegraphics[width=\columnwidth]{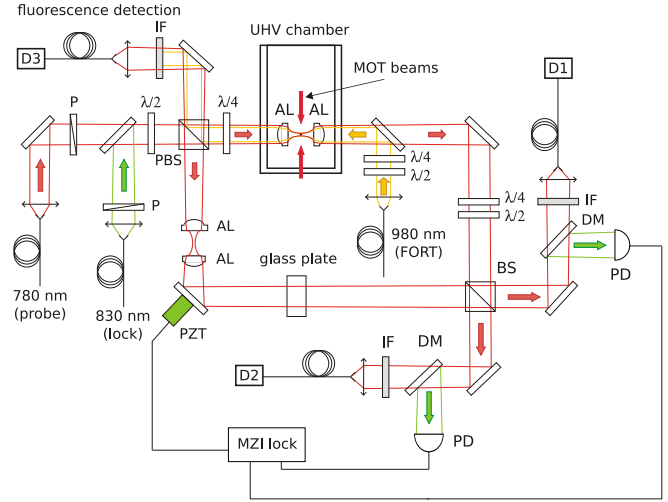}
\caption{\label{setup}
Experimental setup. A single atom located by a far-off-resonant trap (FORT) in
a confocal arrangement of two aspheric lenses (AL) is made part of a
stabilized Mach-Zehnder interferometer. Refer to the text for explanation of
the different components.}
\end{center}
\end{figure}

Figure~\ref{setup} shows a sketch of our experiment. A probe beam is
sent through a stabilized Mach-Zehnder interferometer (MZI). One arm
contains a single $^{87}$Rb atom, trapped at the focus of a confocal aspheric
lens pair (atom arm) in an ultra high vacuum chamber, while the other arm
serves as a phase reference.

The probe is a weak coherent beam with a transverse Gaussian
profile with a waist of $w_{\mathrm{L}}=1.1$\,mm at the focusing lens
($f=4.5$\,mm). During the 
experiment, the frequency of the probe is tuned across the resonance of the 
$5^2{\rm S}_{1/2}, F=2 \rightarrow 5^2{\rm P}_{3/2}, F'=3$ transition of the
D2 line (780\,nm). The ratio of optical power in both arms of
the interferometer is controlled with a half-wave plate and a polarizing
beam-splitter to match at the input ports of the second beam-splitter without
an atom in the focus.
A quarter-wave plate preceeding the focusing lens prepares the probe
into right circular polarization to maximize interaction with the
atom~\cite{our_paper}, before it is focused to a (nominal) waist
$w_{\mathrm{f}}\sim$1.0\,\textmu m. After the lenses, the 
polarization of the probe is converted back to linear with a quarter- and
half-wave plate to match the polarization of the reference arm. The output
modes of the interferometer are then collected into single mode fibers with an
efficiency of $\approx 84\%$ without an atom in the trap, which guide the
light to silicon avalanche photodetectors (APD) D1 and D2. 

The single $^{87}$Rb atom is localized at the focus of the aspheric
lenses by means of a far off-resonant optical dipole trap formed by a tightly
focused light beam at 980\,nm, such that there is 
either one or no atom in the trap at any time due to the `collisional
blockade' mechanism~\cite{schlosser:2002}. Cold atoms are loaded into the
dipole trap from a magneto optical trap (MOT), and the presence of one and only
one atom in the trap is verified by observing strong photon anti-bunching in the
second-order correlation function $g^{(2)}(\tau)$ of the atomic fluorescence.

During the frequency scan of the probe, the atom has a probability to be
excited to $5^2{\rm P}_{3/2}, F=2$ and fall to the $5^2{\rm S}_{1/2}, F=1$
ground state.  
To bring the atom back to the probe transition, light resonant
to the $5^2{\rm S}_{1/2}, F=1 \rightarrow 5^2{\rm P}_{1/2}, F'=2$ transition
of $^{87}$Rb (795\,nm) is sent to the atom together with the
probe beam, and later filtered out with an interference filter IF. 

The phase stability of the interferometer over the measurement time is
ensured by locking it to an off-resonant auxiliary
laser with a wavelength $\lambda=830$\,nm copropagating with the probe.
To ensure that a drift in the frequency of this locking laser does not
change the path length difference significantly, the MZI is adjusted close to
zero path length difference with the help of a glass plate in the reference arm.
This auxiliary light is separated from the probe with dichroic mirrors DM to
provide a feedback signal to a piezoelectric actuator (PZT).

To keep the analysis of the interference pattern simple, we aimed for a
maximal interference contrast in the MZI. Essential for this is a match of
the wavefronts in probe-  and reference arm on the second beam splitter (BS). A
confocal lens pair identical to the one in the probe arm was inserted 
in the reference arm, with an adjustable separation to compensate for any
difference in divergence. The interference contrast
(after coupling into the single mode fibers) had a visibility of
$V=98.0\pm0.2$\,\%.

\section{Phase measurement\label{sec:measurement}}
Once an atom is loaded into the trap (verified by detecting its fluorescence
with detector D3), the MOT beams and quadrupole coil currents are switched off,
and the atom is optically pumped into the $5^2{\rm S}_{1/2}, F=2, m_F=-2
\rightarrow 5^2{\rm P}_{3/2}, F'=3, m_F=-3$ closed cycling transition by
the same probe beam for 20\,ms (see~\cite{our_paper} for details). Then the
detection events at D1 and D2 are 
recorded for 130--140\,ms. After that, the MOT beams are turned on for about
20\,ms to check if the atom is still in the trap. 
If this is the case, the MOT beams are turned off again and the pump, probe and
detection sequence is repeated. Otherwise, the last single probe result is
ignored, and the interferometer outputs are observed 
\emph{without} an atom in the trap for 2\,s with the MOT beams switched off as
a background measurement.

Since our observation is done by detectors probing the light in single
mode optical fibers behind beam splitter, we can express all interference
effects in terms of scalar amplitudes $E$ of field modes in
these fibers, which in the free space part both overlap with the probe and
reference mode.
The optical powers $P_c$ and $P_d$ in the fibers -- in the absence of the atom
and up to a constant -- are given by
\begin{equation}
P_{c,d} = \frac{1}{2}\left[ |E_a|^2 + |E_b|^2 \pm
  2|E_a|\cdot|E_b|\cos\phi_{ab}\right]\,,
\label{eq:emptyinterferometer}
\end{equation}
where $E_a$ and $E_b$ correspond to field amplitudes (with the spatial profile
of the collecting modes) in the atom/reference arms, and $\phi_{ab}$ is the
phase difference between MZI arms. The interferometer has a maximal phase
sensitivity $\partial P/ \partial\phi_{ab}$ for $\phi_{ab} = \pm90^\circ$
where  $|E_a|=|E_b|$. Note that this does not imply equal 
count rates $N_1$ and $N_2$ of the detectors behind the single mode fibers 
due to the different coupling efficiencies in each
channel, and different detector dark count rates. It can be shown that the
locking point with the highest sensitivity for a phase measurement with these
different coupling efficiencies corresponds to count rates
\begin{eqnarray}
N_c^l &= &\frac{N_c^\mathrm{max}-N_c^\mathrm{min}}{2}+B_1\nonumber\\
N_d^l &= &\frac{N_d^\mathrm{max}-N_d^\mathrm{min}}{2}+B_2
\end{eqnarray}
at the output of an empty interferometer, with $N_{c,d}^{\mathrm{min,max}}$
corresponding to the minimal/maximal observed rates for
all phases $\phi_{ab}$, and detector background rates $B_1$ and $B_2$. 

An atom in the trap then scatters photons out of the probe
beam, causing a power drop in the atom arm. With the same convention as in
Eq.~(\ref{eq:emptyinterferometer}), the power levels at the output of the MZI
are given by 
\begin{equation}
P_{c,d}' = \frac{1}{2}\left[ |E_{a}'|^2 + |E_b|^2 \pm
  2|E_a'|\cdot|E_b|\cos\phi'_{ab}\right]\,,
\end{equation}
where $|E_b|$ remains unchanged, and the primes indicate changed values in the
atom arm. The phase difference between the arms is given by 
\begin{equation}
\phi'_{ab} = \arccos \frac{P_c'-P_d'}{\left(P_c +
      P_d\right)\sqrt{T}}\,,
\label{eq:phase1}
\end{equation}
where $T$ is the transmission of the probe beam in the atom arm,
\begin{equation}
T =
\left|\frac{E_a'}{E_a}\right|^2=\frac{2\left(P_c'+P_d'\right)}{P_c+P_d}-1\,. 
\label{eq:phase2}
\end{equation}
Note that for the relations in Eq. (\ref{eq:phase1}) and (\ref{eq:phase2}) to
hold, $|E_a|=|E_b|$, which we verified by the high visibility of the empty
interferometer. The actual phase shift induced by the atom is then simply
\begin{equation}
\delta \phi = \phi'_{ab} - \phi_{ab}\,.
\end{equation}

In the same experimental run (i.e., for the same detuning of the probe
frequency), we have also performed an independent measurement of the
transmission $T$ of the probe beam with the reference arm blocked using the
same measurement sequence, which leads to a better signal/noise ratio.

\section{Theory\label{theory}}
The electric field at the input of the beam splitter $\vec{E}_a'(\vec{r})$
results from the superposition of the field of the probe
$\vec{E}_{a}(\vec{r})$ with the field scattered by the atom $\vec{E}_{\mathrm{sc}}(\vec{r})$: 
 \begin{equation}
\vec{E}_a'(\vec{r})=\vec{E}_{a}(\vec{r})+\vec{E}_{\mathrm{sc}}(\vec{r})
\end{equation}
The spatial dependency of the scattered field $\vec{E}_{sc}(\vec{r})$ is that
of a rotating electrical dipole, with an amplitude proportional to the
exciting electrical field amplitude $E_A$ at the location of the atom.
Far away from the dipole ($r\gg\lambda$), it takes the
form~\cite{zumofen:2008,mk:2009} 
\begin{equation}
  \vec{E}_{\mathrm{sc}}(\vec{r}) = \frac{3E_Ae^{i(kr+\pi/2)}}{2kr}
  \left[\hat{\epsilon}_+-(\hat{\epsilon}_+\cdot\hat{r})\hat{r}\right]
  \frac{i\Gamma}{2\Delta+i\Gamma}\,,
\end{equation}
where $\epsilon_+$ is the unit vector of circular polarization.
The frequency-dependent phase enters via the Lorentzian function ($\Delta$ is
the detuning from resonance, $\Gamma$ the natural linewidth of the atomic
transition). The $\pi/2$ phase reflects the {\it lag} of the atom response
with respect to the excitation field $E_{A}$ by $\pi/2$ on resonance.


The superposition of the probe and atomic response leads to an amplitude
$E_a'$ in the collection mode.
Following~\cite{mk:2009}, we assume that the collection and probe mode
coincide in the absence of the atom, $\vec{G}_{a}(\vec{r})\propto
\vec{E}_{a}(\vec{r})$. With the normalization
$\int\left[\vec{E}_{a}(\vec{r})\cdot\vec{G}_a^*(\vec{r})\right]dS=E_a$, where
$dS$ is an element of the integration surface parallel to the local wavefront
of the probe mode somewhere after the atom, $E'_a$ is given by 
\begin{equation}
  E_a'=\int\left[\left(\vec{E}_{a}(\vec{r})+\vec{E}_{sc}(\vec{r})\right)\cdot\vec{G}_a^*(\vec{r})\right]dS\,.
\end{equation}

Phase
shift and transmission of the probe beam are only determined by the complex
ratio $E_a'/E_a$. The extension of the result for Gaussian mode profiles
presented in~\cite{mk:2009} with the Lorentzian term leads to
\begin{equation}
{E_a'\over E_a} =
1-\frac{R_{\mathrm{sc}}}{2}\frac{i\Gamma}{2\Delta+i\Gamma}\,,
\label{eq:aratio}
\end{equation}
where $R_{\mathrm{sc}}$ is the scattering ratio for the
probe which depends only on a focusing strength $u:=w_L/f$ of the Gaussian
beam. The atom-induced phase shift of the probe mode is then given by
\begin{equation}
\label{phaseshift_final}
\delta \phi=\arg(E_a'/E_a)\,.
\end{equation}

\begin{figure}
\begin{center}
\includegraphics[width=1.0\columnwidth]{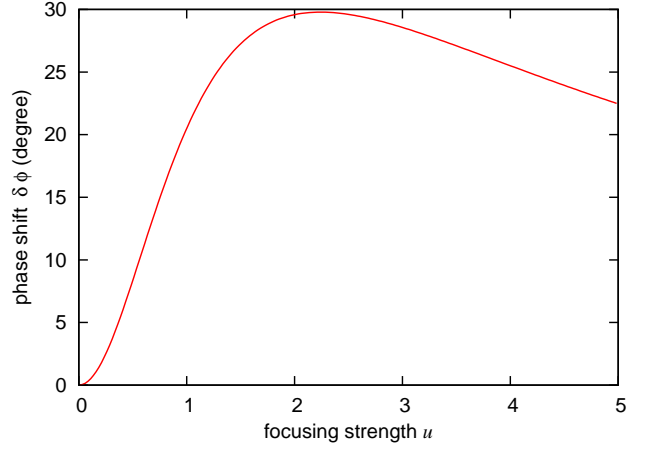}
\caption{\label{phase_theory}
Phase shift $\delta\phi$ of a beam with Gaussian profile and focusing strength
$u$ (as defined in the text) due to a single atom at a detuning
$\Delta=-\Gamma/2$ from resonance. A maximal phase shift of $29.78^\circ$ is
expected for $u=2.24$.}
\end{center}
\end{figure}

Figure~\ref{phase_theory} shows the dependence of the expected phase shift
on the focusing strength $u$. The maximal phase shift is experienced for
$\Delta=-\Gamma/2$, and reaches about $30^\circ$ for this `fiber-atom-fiber'
interface at $u=2.24$.
Our experimental parameters correspond to $u=0.244$ or $R_{\mathrm{sc}}=0.16$,
so we expect a maximal phase shift of $2.3^\circ$ at detuning $\Delta=\Gamma/2$.

\section{Results and Discussion}\label{sec:results}
\begin{figure}
\begin{center}
\includegraphics[width=1.0\columnwidth]{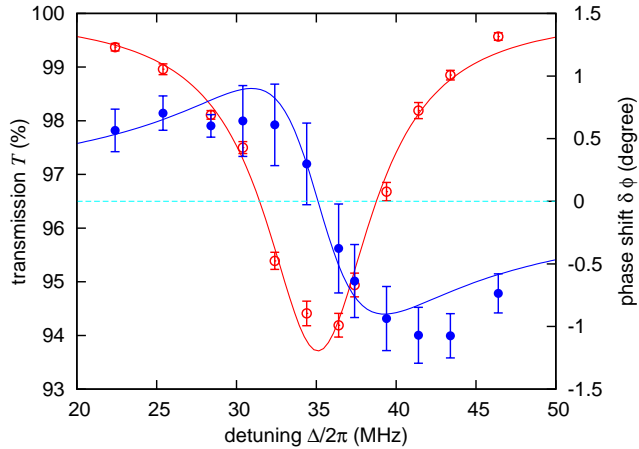}
\caption{\label{plot}
Phase shift $\delta\phi$ observed on a weak coherent probe beam tuned across
the resonance of a single atom (filled symbols), showing the dispersive
character from phase retardation below resonance to phase advancement above
resonance.  The transmission $T$ of the same probe is shown for reference
(open circles). Solid lines correspond to theoretical values (see text for
details). 
} 
\end{center}
\end{figure}

Figure~\ref{plot} shows the experimentally observed phase shift and
transmission of the probe beam as a function of detuning from the natural
resonant frequency. Our transmission results can be modeled by the
expression obtained from Eq.~(\ref{eq:aratio}), 
\begin{equation}
T = \left|\frac{E_a'}{E_a}\right|^2 =
1-\frac{\Gamma^2R_{\mathrm{sc}}(1-R_{\mathrm{sc}}/4)}{4(\Delta-\Delta_0)^2+\Gamma^2}\,,
\end{equation}
with fit parameters $\Gamma/2\pi=8.20\pm0.47$\,MHz,
$\Delta_0/2\pi=35.1\pm0.2$\,MHz, and $R_{\mathrm{sc}}=0.064\pm0.004$. The
latter is not only governed by the 
focusing parameter, but also experimental uncertainties about the exact field in
the focus and the atomic position, while $\Delta_0$ reflects the trap-induced
AC Stark shift. The transmission linewidth $\Gamma$ slightly exceeds 
the natural linewidth $\Gamma_{\mathrm{nat}}/2\pi=6$\,MHz of the atomic
transition. One reason for this is the finite linewidth of the probe laser,
measured as $\Delta\nu_{\mathrm{L}}=750$\,kHz FWHM. Other contribution is
Doppler broadening and a position-dependent detuning due to residual
motion of the atom in the trap.

The solid line shown together with the phase shift results in
Fig.~\ref{plot} corresponds to Eq.~(\ref{phaseshift_final}), with the
parameters $\Gamma$, $\Delta_0$, and $R_{\mathrm{sc}}$ from the transmission
fit, in good agreement with the experimental values. As expected, above the
atomic resonance an advance of the phase is observed, while below resonance
the atom introduces a phase lag to the probe beam.

The maximal phase shift of $0.97^\circ$  according to
Eq.~(\ref{phaseshift_final}) and the fit parameters from the transmission
measurement at $\Delta=\Gamma/2$ is about 2.6 times smaller than what we
would expect for our focusing parameter. We amount this discrepancy to two
contributions: firstly, the lenses in the experiment are not ideal, so the
calculated value of $R_{\mathrm{sc}}$ may not reflect the actual field strength at the
atom. An independent measurement of the field at the focus \cite{wilson:1997,
  quabis:2001, rhodes:2002}
would help to assess this contribution quantitatively.
Secondly, the atom in the trap is not stationary, thus the average probe field
strength that it experiences is lower than the calculated value in the focal
position. With our trap frequencies of $\nu_t\approx70$ kHz 
in transverse and $\nu_z\approx 20$ kHz in longitudinal direction together
with the estimated temperature of the atom of $\approx100\,\mu$K
(as measured in similar trap configurations \cite{weber:2006, tuchendler:2008}),
the atom has a position uncertainty of $\sigma_t\approx 220$\,nm and $\sigma_z\approx780$\,nm, respectively, reducing
the scattering ratio $R_{\mathrm{sc}}$ by 23\% \cite{mk:2009}. However, the
scattering ratio is very sensitive to temperature of the atom, and doubling of
the temperature alone would explain the discrepancy between
theory and experiment. Additional cooling techniques
\cite{tuchendler:2008, chu:1992, chu:1996} would help to reduce this contribution.

\section{Conclusion}
In summary, we have measured the phase shift that the presence of a single
$^{87}$Rb atom imposes on a near resonant focused light field. The theoretical
model suggests that realistic experimental improvement in the focusing strength
and on the atom localization to levels comparable to what is achieved in ion
traps focusing quality will lead to substantial phase shifts on a light beam
by a single atom. With a control of the atomic state by another photon, this
atom-light interface may form relatively simple building block in a phase
gate between photonic qubits.

\section{Acknowledgment}
We acknowledge the support of this work by the National Research Foundation \&
Ministry of Education in Singapore.


\begin{thebibliography}{22}
\expandafter\ifx\csname natexlab\endcsname\relax\def\natexlab#1{#1}\fi
\expandafter\ifx\csname bibnamefont\endcsname\relax
  \def\bibnamefont#1{#1}\fi
\expandafter\ifx\csname bibfnamefont\endcsname\relax
  \def\bibfnamefont#1{#1}\fi
\expandafter\ifx\csname citenamefont\endcsname\relax
  \def\citenamefont#1{#1}\fi
\expandafter\ifx\csname url\endcsname\relax
  \def\url#1{\texttt{#1}}\fi
\expandafter\ifx\csname urlprefix\endcsname\relax\def\urlprefix{URL }\fi
\providecommand{\bibinfo}[2]{#2}
\providecommand{\eprint}[2][]{\url{#2}}

\bibitem[{\citenamefont{Cirac et~al.}(1997)\citenamefont{Cirac, Zoller, Kimble,
  and Mabuchi}}]{cirac:1997}
\bibinfo{author}{\bibfnamefont{J.~I.} \bibnamefont{Cirac}},
  \bibinfo{author}{\bibfnamefont{P.}~\bibnamefont{Zoller}},
  \bibinfo{author}{\bibfnamefont{H.~J.} \bibnamefont{Kimble}},
  \bibnamefont{and} \bibinfo{author}{\bibfnamefont{H.}~\bibnamefont{Mabuchi}},
  \bibinfo{journal}{Phys.\ Rev.\ Lett.} \textbf{\bibinfo{volume}{78}},
  \bibinfo{pages}{3221} (\bibinfo{year}{1997}).

\bibitem[{\citenamefont{Duan et~al.}(2001)\citenamefont{Duan, Lukin, Cirac, and
  Zoller}}]{duan:2001}
\bibinfo{author}{\bibfnamefont{L.-M.} \bibnamefont{Duan}},
  \bibinfo{author}{\bibfnamefont{M.~D.} \bibnamefont{Lukin}},
  \bibinfo{author}{\bibfnamefont{J.~I.} \bibnamefont{Cirac}}, \bibnamefont{and}
  \bibinfo{author}{\bibfnamefont{P.}~\bibnamefont{Zoller}},
  \bibinfo{journal}{Nature} \textbf{\bibinfo{volume}{414}},
  \bibinfo{pages}{413} (\bibinfo{year}{2001}).

\bibitem[{\citenamefont{Pinkse et~al.}(2000)\citenamefont{Pinkse, Fischer,
  Mauncz, and Rempe}}]{rempe:2000}
\bibinfo{author}{\bibfnamefont{P.~W.~H.} \bibnamefont{Pinkse}},
  \bibinfo{author}{\bibfnamefont{T.}~\bibnamefont{Fischer}},
  \bibinfo{author}{\bibfnamefont{P.}~\bibnamefont{Mauncz}}, \bibnamefont{and}
  \bibinfo{author}{\bibfnamefont{G.}~\bibnamefont{Rempe}},
  \bibinfo{journal}{Nature} \textbf{\bibinfo{volume}{404}}
  (\bibinfo{year}{2000}).

\bibitem[{\citenamefont{Boozer et~al.}(2007)\citenamefont{Boozer, Boca, Miller,
  Northup, and Kimle}}]{kimble:2007}
\bibinfo{author}{\bibfnamefont{A.~D.} \bibnamefont{Boozer}},
  \bibinfo{author}{\bibfnamefont{A.}~\bibnamefont{Boca}},
  \bibinfo{author}{\bibfnamefont{R.}~\bibnamefont{Miller}},
  \bibinfo{author}{\bibfnamefont{T.~E.} \bibnamefont{Northup}},
  \bibnamefont{and} \bibinfo{author}{\bibfnamefont{H.~J.} \bibnamefont{Kimle}},
  \bibinfo{journal}{Phys.\ Rev.\ Lett.} \textbf{\bibinfo{volume}{98}}
  (\bibinfo{year}{2007}).

\bibitem[{\citenamefont{Tey et~al.}(2009)\citenamefont{Tey, Maslennikov, Liew,
  Aljunid, Huber, Chng, Chen, Scarani, and Kurtsiefer}}]{mk:2009}
\bibinfo{author}{\bibfnamefont{M.~K.} \bibnamefont{Tey}},
  \bibinfo{author}{\bibfnamefont{G.}~\bibnamefont{Maslennikov}},
  \bibinfo{author}{\bibfnamefont{T.~C.~H.} \bibnamefont{Liew}},
  \bibinfo{author}{\bibfnamefont{S.~A.} \bibnamefont{Aljunid}},
  \bibinfo{author}{\bibfnamefont{F.}~\bibnamefont{Huber}},
  \bibinfo{author}{\bibfnamefont{B.}~\bibnamefont{Chng}},
  \bibinfo{author}{\bibfnamefont{Z.}~\bibnamefont{Chen}},
  \bibinfo{author}{\bibfnamefont{V.}~\bibnamefont{Scarani}}, \bibnamefont{and}
  \bibinfo{author}{\bibfnamefont{C.}~\bibnamefont{Kurtsiefer}},
  \bibinfo{journal}{New J. Phys.} \textbf{\bibinfo{volume}{11}},
  \bibinfo{pages}{043011} (\bibinfo{year}{2009}).

\bibitem[{\citenamefont{Zumofen et~al.}(2008)\citenamefont{Zumofen, Mojarad,
  Sandoghdar, and Agio}}]{zumofen:2008}
\bibinfo{author}{\bibfnamefont{G.}~\bibnamefont{Zumofen}},
  \bibinfo{author}{\bibfnamefont{N.}~\bibnamefont{Mojarad}},
  \bibinfo{author}{\bibfnamefont{V.}~\bibnamefont{Sandoghdar}},
  \bibnamefont{and} \bibinfo{author}{\bibfnamefont{M.}~\bibnamefont{Agio}},
  \bibinfo{journal}{Phys. Rev. Lett.} \textbf{\bibinfo{volume}{101}},
  \bibinfo{pages}{180404} (\bibinfo{year}{2008}).

\bibitem[{\citenamefont{Vamivakas et~al.}(2007)\citenamefont{Vamivakas,
  Atature, Dreiser, Yilmaz, Badolato, Swan, Goldberg, Imamoglu, and
  Unlu}}]{imamoglu:2007}
\bibinfo{author}{\bibfnamefont{A.~N.} \bibnamefont{Vamivakas}},
  \bibinfo{author}{\bibfnamefont{M.}~\bibnamefont{Atature}},
  \bibinfo{author}{\bibfnamefont{J.}~\bibnamefont{Dreiser}},
  \bibinfo{author}{\bibfnamefont{S.~T.} \bibnamefont{Yilmaz}},
  \bibinfo{author}{\bibfnamefont{A.}~\bibnamefont{Badolato}},
  \bibinfo{author}{\bibfnamefont{A.~K.} \bibnamefont{Swan}},
  \bibinfo{author}{\bibfnamefont{B.~B.} \bibnamefont{Goldberg}},
  \bibinfo{author}{\bibfnamefont{A.}~\bibnamefont{Imamoglu}}, \bibnamefont{and}
  \bibinfo{author}{\bibfnamefont{M.~S.} \bibnamefont{Unlu}},
  \bibinfo{journal}{Nano Letters} \textbf{\bibinfo{volume}{7}}
  (\bibinfo{year}{2007}).

\bibitem[{\citenamefont{Wrigge et~al.}(2008)\citenamefont{Wrigge, Gerhardt,
  Hwang, Zumofen, and Sangoghdar}}]{wrigge:2008}
\bibinfo{author}{\bibfnamefont{G.}~\bibnamefont{Wrigge}},
  \bibinfo{author}{\bibfnamefont{I.}~\bibnamefont{Gerhardt}},
  \bibinfo{author}{\bibfnamefont{J.}~\bibnamefont{Hwang}},
  \bibinfo{author}{\bibfnamefont{G.}~\bibnamefont{Zumofen}}, \bibnamefont{and}
  \bibinfo{author}{\bibfnamefont{V.}~\bibnamefont{Sangoghdar}},
  \bibinfo{journal}{Nature Physics} \textbf{\bibinfo{volume}{4}}
  (\bibinfo{year}{2008}).

\bibitem[{\citenamefont{Tey et~al.}(2008)\citenamefont{Tey, Chen, Aljunid,
  Chng, Huber, Maslennikov, and Kurtsiefer}}]{our_paper}
\bibinfo{author}{\bibfnamefont{M.~K.} \bibnamefont{Tey}},
  \bibinfo{author}{\bibfnamefont{Z.}~\bibnamefont{Chen}},
  \bibinfo{author}{\bibfnamefont{S.~A.} \bibnamefont{Aljunid}},
  \bibinfo{author}{\bibfnamefont{B.}~\bibnamefont{Chng}},
  \bibinfo{author}{\bibfnamefont{F.}~\bibnamefont{Huber}},
  \bibinfo{author}{\bibfnamefont{G.}~\bibnamefont{Maslennikov}},
  \bibnamefont{and}
  \bibinfo{author}{\bibfnamefont{C.}~\bibnamefont{Kurtsiefer}},
  \bibinfo{journal}{Nature Physics} \textbf{\bibinfo{volume}{4}}
  (\bibinfo{year}{2008}).

\bibitem[{\citenamefont{Savage et~al.}(1990)\citenamefont{Savage, Braunstein,
  and Walls}}]{walls:1990}
\bibinfo{author}{\bibfnamefont{C.~M.} \bibnamefont{Savage}},
  \bibinfo{author}{\bibfnamefont{S.~L.} \bibnamefont{Braunstein}},
  \bibnamefont{and} \bibinfo{author}{\bibfnamefont{D.~F.} \bibnamefont{Walls}},
  \bibinfo{journal}{Opt. Lett.} \textbf{\bibinfo{volume}{15}},
  \bibinfo{pages}{628} (\bibinfo{year}{1990}).

\bibitem[{\citenamefont{Turchette et~al.}(1995)\citenamefont{Turchette, Hood,
  Lange, Mabuchi, and Kimble}}]{turchette:1995}
\bibinfo{author}{\bibfnamefont{Q.~A.} \bibnamefont{Turchette}},
  \bibinfo{author}{\bibfnamefont{C.~J.} \bibnamefont{Hood}},
  \bibinfo{author}{\bibfnamefont{W.}~\bibnamefont{Lange}},
  \bibinfo{author}{\bibfnamefont{H.}~\bibnamefont{Mabuchi}}, \bibnamefont{and}
  \bibinfo{author}{\bibfnamefont{H.~J.} \bibnamefont{Kimble}},
  \bibinfo{journal}{Phys. Rev. Lett.} \textbf{\bibinfo{volume}{75}},
  \bibinfo{pages}{4710} (\bibinfo{year}{1995}).

\bibitem[{\citenamefont{Fushman et~al.}(2008)\citenamefont{Fushman, Englund,
  Faraon, Stoltz, Petroff, and Vuckovic}}]{fushman:2008}
\bibinfo{author}{\bibfnamefont{I.}~\bibnamefont{Fushman}},
  \bibinfo{author}{\bibfnamefont{D.}~\bibnamefont{Englund}},
  \bibinfo{author}{\bibfnamefont{A.}~\bibnamefont{Faraon}},
  \bibinfo{author}{\bibfnamefont{N.}~\bibnamefont{Stoltz}},
  \bibinfo{author}{\bibfnamefont{P.}~\bibnamefont{Petroff}}, \bibnamefont{and}
  \bibinfo{author}{\bibfnamefont{J.}~\bibnamefont{Vuckovic}},
  \bibinfo{journal}{Science} \textbf{\bibinfo{volume}{320}},
  \bibinfo{pages}{769} (\bibinfo{year}{2008}).

\bibitem[{\citenamefont{Zibrov et~al.}(1996)\citenamefont{Zibrov, Lukin,
  Hollberg, Nikonov, Scully, Robinson, and Velichansky}}]{zibrov:1996}
\bibinfo{author}{\bibfnamefont{A.~S.} \bibnamefont{Zibrov}},
  \bibinfo{author}{\bibfnamefont{M.~D.} \bibnamefont{Lukin}},
  \bibinfo{author}{\bibfnamefont{L.}~\bibnamefont{Hollberg}},
  \bibinfo{author}{\bibfnamefont{D.~E.} \bibnamefont{Nikonov}},
  \bibinfo{author}{\bibfnamefont{M.~O.} \bibnamefont{Scully}},
  \bibinfo{author}{\bibfnamefont{H.~G.} \bibnamefont{Robinson}},
  \bibnamefont{and} \bibinfo{author}{\bibfnamefont{V.~L.}
  \bibnamefont{Velichansky}}, \bibinfo{journal}{Phys. Rev. Lett.}
  \textbf{\bibinfo{volume}{76}}, \bibinfo{pages}{3935} (\bibinfo{year}{1996}).

\bibitem[{\citenamefont{Zumofen et~al.}(2009)\citenamefont{Zumofen, Mojarad,
  and Agio}}]{zumofen:2009}
\bibinfo{author}{\bibfnamefont{G.}~\bibnamefont{Zumofen}},
  \bibinfo{author}{\bibfnamefont{N.}~\bibnamefont{Mojarad}}, \bibnamefont{and}
  \bibinfo{author}{\bibfnamefont{M.}~\bibnamefont{Agio}},
  \bibinfo{journal}{Nuovo Cimento C} \textbf{\bibinfo{volume}{31}},
  \bibinfo{pages}{475} (\bibinfo{year}{2009}).

\bibitem[{\citenamefont{Schlosser et~al.}(2002)\citenamefont{Schlosser,
  Reymond, and Grangier}}]{schlosser:2002}
\bibinfo{author}{\bibfnamefont{N.}~\bibnamefont{Schlosser}},
  \bibinfo{author}{\bibfnamefont{G.}~\bibnamefont{Reymond}}, \bibnamefont{and}
  \bibinfo{author}{\bibfnamefont{P.}~\bibnamefont{Grangier}},
  \bibinfo{journal}{Phys. Rev. Lett.} \textbf{\bibinfo{volume}{89}},
  \bibinfo{pages}{023005} (\bibinfo{year}{2002}).

\bibitem[{\citenamefont{Juskaitis and Wilson}(1997)}]{wilson:1997}
\bibinfo{author}{\bibfnamefont{R.}~\bibnamefont{Juskaitis}} \bibnamefont{and}
  \bibinfo{author}{\bibfnamefont{T.}~\bibnamefont{Wilson}},
  \bibinfo{journal}{Journal of Microscopy} \textbf{\bibinfo{volume}{189}},
  \bibinfo{pages}{8} (\bibinfo{year}{1997}).

\bibitem[{\citenamefont{Quabis et~al.}(2001)\citenamefont{Quabis, Dorn,
  Eberler, O., and Leuchs}}]{quabis:2001}
\bibinfo{author}{\bibfnamefont{S.}~\bibnamefont{Quabis}},
  \bibinfo{author}{\bibfnamefont{R.}~\bibnamefont{Dorn}},
  \bibinfo{author}{\bibfnamefont{M.}~\bibnamefont{Eberler}},
  \bibinfo{author}{\bibfnamefont{G.}~\bibnamefont{O.}}, \bibnamefont{and}
  \bibinfo{author}{\bibfnamefont{G.}~\bibnamefont{Leuchs}},
  \bibinfo{journal}{Applied Physics B} \textbf{\bibinfo{volume}{72}},
  \bibinfo{pages}{109} (\bibinfo{year}{2001}).

\bibitem[{\citenamefont{Rhodes et~al.}(2002)\citenamefont{Rhodes, Nugent, and
  Roberts}}]{rhodes:2002}
\bibinfo{author}{\bibfnamefont{S.~K.} \bibnamefont{Rhodes}},
  \bibinfo{author}{\bibfnamefont{K.~A.} \bibnamefont{Nugent}},
  \bibnamefont{and} \bibinfo{author}{\bibfnamefont{A.}~\bibnamefont{Roberts}},
  \bibinfo{journal}{J. Opt. Soc. Am. A} \textbf{\bibinfo{volume}{19}},
  \bibinfo{pages}{1689} (\bibinfo{year}{2002}).

\bibitem[{\citenamefont{Weber et~al.}(2006)\citenamefont{Weber, Volz, Saucke,
  Kurtsiefer, and Weinfurter}}]{weber:2006}
\bibinfo{author}{\bibfnamefont{M.}~\bibnamefont{Weber}},
  \bibinfo{author}{\bibfnamefont{J.}~\bibnamefont{Volz}},
  \bibinfo{author}{\bibfnamefont{K.}~\bibnamefont{Saucke}},
  \bibinfo{author}{\bibfnamefont{C.}~\bibnamefont{Kurtsiefer}},
  \bibnamefont{and}
  \bibinfo{author}{\bibfnamefont{H.}~\bibnamefont{Weinfurter}},
  \bibinfo{journal}{Physical Review A} \textbf{\bibinfo{volume}{73}},
  \bibinfo{pages}{043406} (\bibinfo{year}{2006}).

\bibitem[{\citenamefont{Tuchendler et~al.}(2008)\citenamefont{Tuchendler,
  Lance, Browaeys, Sortais, and Grangier}}]{tuchendler:2008}
\bibinfo{author}{\bibfnamefont{C.}~\bibnamefont{Tuchendler}},
  \bibinfo{author}{\bibfnamefont{A.~M.} \bibnamefont{Lance}},
  \bibinfo{author}{\bibfnamefont{A.}~\bibnamefont{Browaeys}},
  \bibinfo{author}{\bibfnamefont{Y.~R.~P.} \bibnamefont{Sortais}},
  \bibnamefont{and} \bibinfo{author}{\bibfnamefont{P.}~\bibnamefont{Grangier}},
  \bibinfo{journal}{Physical Review A} \textbf{\bibinfo{volume}{78}},
  \bibinfo{pages}{033425} (\bibinfo{year}{2008}).

\bibitem[{\citenamefont{Kasevich and Chu}(1992)}]{chu:1992}
\bibinfo{author}{\bibfnamefont{M.}~\bibnamefont{Kasevich}} \bibnamefont{and}
  \bibinfo{author}{\bibfnamefont{S.}~\bibnamefont{Chu}},
  \bibinfo{journal}{Phys. Rev. Lett.} \textbf{\bibinfo{volume}{69}},
  \bibinfo{pages}{1741} (\bibinfo{year}{1992}).

\bibitem[{\citenamefont{Lee et~al.}(1996)\citenamefont{Lee, Adams, Kasevich,
  and Chu}}]{chu:1996}
\bibinfo{author}{\bibfnamefont{H.~J.} \bibnamefont{Lee}},
  \bibinfo{author}{\bibfnamefont{C.~S.} \bibnamefont{Adams}},
  \bibinfo{author}{\bibfnamefont{M.}~\bibnamefont{Kasevich}}, \bibnamefont{and}
  \bibinfo{author}{\bibfnamefont{S.}~\bibnamefont{Chu}},
  \bibinfo{journal}{Phys. Rev. Lett.} \textbf{\bibinfo{volume}{76}},
  \bibinfo{pages}{2658} (\bibinfo{year}{1996}).

\end{thebibliography}
\end{document}